\documentclass[a4paper,twocolumn,superscriptaddress]{revtex4-1}
\usepackage{amsfonts}
\usepackage{natbib}
\usepackage{caption}
\usepackage{subcaption}
\usepackage{listings}
\usepackage{bm}
\usepackage{appendix}
\usepackage{amsmath,amssymb,graphicx}
\usepackage{lipsum}
\usepackage[percent]{overpic}

\begin{document}

\title{Skew ray tracing in a step-index optical fiber using Geometric Algebra}

\author{Angeleene S. Ang}
\email{angeleene.ang@gmail.com}
\affiliation{Ateneo de Manila University, Department of Physics, 
  Loyola Heights, Quezon City, Philippines 1108}

\author{Quirino M. Sugon, Jr.}
\affiliation{Manila Observatory, Upper Atmosphere Division, 
  Ateneo de Manila University Campus}
\affiliation{Ateneo de Manila University, Department of Physics, 
  Loyola Heights, Quezon City, Philippines 1108}

\author{Daniel J. McNamara}
\affiliation{Manila Observatory, Upper Atmosphere Division, 
  Ateneo de Manila University Campus}
\affiliation{Ateneo de Manila University, Department of Physics, 
  Loyola Heights, Quezon City, Philippines 1108}

\begin{abstract}

We used Geometric Algebra to compute the paths of skew rays in a cylindrical, step-index multimode optical fiber. 
To do this, we used the vector addition form for the law of propagation, the exponential of an imaginary vector form for the law of refraction, and the juxtaposed vector product form for the law of reflection. 
In particular, the exponential forms of the vector rotations enables us to take advantage of the addition or subtraction of exponential arguments of two rotated vectors in the derivation of the ray tracing invariants in cylindrical and spherical coordinates. 
We showed that the light rays inside the optical fiber trace a polygonal helical path characterized by three invariants that relate successive reflections inside the fiber: the ray path distance, the difference in axial distances, and the difference in the azimuthal angles. 
We also rederived the known generalized formula for the numerical aperture for skew rays, which simplifies to the standard form for meridional rays.

\end{abstract}

\date{\today}

\maketitle

\section{Introduction} \label{sec:intro}







Optical fibers are waveguides used for optical communication as first illustrated by Tyndall in 1870\cite{Kapany}. 
Single-mode fibers have radii of about 8 microns, while multimode fibers have about 60 microns\cite{opticalswitching}. 
The description of light propagation in single-mode fibers require wave optics, while that in multimode fibers, the ray approximation is sufficient.\cite{lovesnyder} 
In this paper, we shall only talk of multimode fibers using ray or geometric optics, i.e, no diffraction, interference, or polarization.

Optical fibers can be classified depending on the functional dependence of the refractive index on the fiber radius: step-index or graded-index (GRIN). 
Step-index can be multi-step\cite{zubia} while, graded-index fibers can have refractive index functions that are parabolic. 
Here, we shall focus only on step-index fibers, where the refractive index of the core is constant and light rays are guided through total internal reflection. 

If the rays lie only on a single plane, the rays are said to be meridional; if the rays trace a discrete helix, the rays are skew.
Most textbooks describe only the meridional rays and mention skew rays only in passing. 
The reason for this is that the mathematics used to describe skew rays is difficult. 
\cite{leviBook,handbook,iizukaBook}

The study of geometric optics ray tracing for skew rays is still an important problem for two reasons. 
The first is pedagogical: skew ray tracing for laser light in large fibers, e.g., HeNe laser in 1 cm diameter glass fiber, can be used to verify the helical properties of skew rays in a student laboratory. 
The second is practical: skew rays tracing equations can serve as the starting point for the analysis of electric fields at the fiber walls\cite{Lin} and can provide the basis of ray tracing algorithms for optical fibers\cite{Witkowski}, which may even include the Goos-H\"{a}nchen effect for reflection\cite{kovac2}. 

To construct a ray tracing algorithm, we need to describe rays mathematically. We can do this in three ways.

The first way is by labeling points, lines, and angles, then use geometry and trigonometry to determine the scalar invariants, e.g., the axial path length between reflections, axial angle of incident and reflected rays, and the numerical aperture. 
But the system for naming these geometrical quantities vary from author to author, making it difficult to compare results.
\cite{Potter, senior,kovac,sugonSPP, lovesnyder}
To solve the problem of the angle naming conventions, we define the directions of the vectors in terms of the polar angles $\theta$ and the azimuthal angle $\phi$ in spherical coordinates, distinguished only by their subscripts, e.g. $\theta_{\sigma}$ and $\phi_{\sigma}$ for the incident ray $\bm \sigma$ and $\phi_{\eta}$ for the normal vector $\bm \eta$.

The second way is to use vectors for ray tracing. 
In general, we need five vectors: the initial position of the ray $\mathbf r$, the final position of the ray $\mathbf r'$, the propagation vector $\bm \sigma$, the reflected vector $\bm \sigma'$, and the vector normal to the surface $\bm \eta'$. 
The task of ray tracing then is to relate these variables using a set of equations, such as those given in Klein and Furtak\cite{kleitak}, though in a slightly different form involving the concavity function $c_{\sigma \eta k} = \pm 1$:
\begin{subequations}
\label{eq:kleitakraytracing}
\begin{align}
  \mathbf r_{k+1} &= 
  \mathbf r_k + s_{k+1} \, \bm \sigma_{k+1} ,
\\
\label{eq:kleitakreflection}
\bm \sigma_k' &= \bm \sigma_k - 2c_{\sigma \eta k} \bm \eta_k \cos \beta_k ,
\\
n_{k+1} \bm \sigma_{k+1} &= n_k \bm \sigma_k \notag
\\ &
\label{eq:kleitakrefraction}
+ c_{\sigma \eta k}(n_{k+1} \cos \beta_{k+1} - n_k \cos \beta_k)\bm \eta_k ,
\end{align}
\end{subequations}
which corresponds to propagation, reflection, and refraction. 
But these ray tracing equations are difficult to apply if we wish to compute for analytic solutions, such as reflected ray and final position after a certain number of reflections within the fiber. That is why ray tracing algorithms do not use analytical solutions but rather iterative computations using the equations described\cite{Witkowski, kovac2}.

The third way is to use vectors in ray tracing but within the framework of the Eikonal method.
\cite{gbur,bornwolf}
Here, the position $\mathbf R$ of the ray satisfies a vector differential equation:
\begin{equation}
\frac{\mathrm{d}}{\mathrm{d}s}\left(n\frac{\mathrm{d}\mathbf{R}}{\mathrm{d}s}\right) = \nabla n ,
\end{equation}
where $s$ is the propagation distance and $n$ is the refractive index. This method lends itself readily to the analysis of invariants in step-index fibers as shown by Zubia et al. \cite{zubia}. This method may also be applied to graded-index fibers using computational methods
\cite{Witkowski,Sakamoto,Southwell}. 
This differential equation approach may be powerful and lends itself well to computational methods, but it does not exploit the simple vector geometry of rays for step-index fibers. 




%
%
%

In this paper, we shall not use the Eikonal method nor the reflection and refraction equations in Eqs.~(\ref{eq:kleitakreflection}) and (\ref{eq:kleitakrefraction}). Instead, we shall construct the ray tracing equations using the exponential rotational operators and direct vector products in Geometric Algebra: 
\begin{subequations}
\begin{align}
\label{eq:pauliintro}
\mathbf a \mathbf b &= \mathbf a \cdot \mathbf b + i \, \mathbf a \times \mathbf b ,
\\
\label{eq:eulerintro}
e^{i \mathbf n \theta} &= \cos \theta + i\mathbf n \sin \theta ,
\end{align}
\end{subequations}
which are the Pauli identity\cite{baylis1,lounesto} and Euler's Theorem, respectively.
For example, the law of reflection and refraction can be expressed as
\begin{subequations}
\begin{align}
\label{eq:simplereflectionintro}
      \bm \sigma_{k+1} &= 
    - \bm \eta_k \bm \sigma_k \bm \eta_k ,
  \\
  \label{eq:refractionintro}
     \bm \sigma_{k+1} &= 
  \bm \sigma_k \, e^{i c_{\sigma \eta k} \mathbf e_{\sigma \eta k} (\beta_k - \beta_{k+1} )} .
\end{align}
\end{subequations}

The direct vector product form of the law of reflection\cite{snygg, lounesto, doranBook} in Eq.~(\ref{eq:simplereflectionintro}) and the exponential rotation form of the law of refraction\cite{sugonAJP}  in Eq.~(\ref{eq:refractionintro}) were already known before. 
These laws and their variants were used in geometric optics for the derivation of ray tracing equations for spherical lenses and mirrors for different ray cases: finite skew, paraxial skew, finite meridional, and paraxial meridional\cite{sugonBook,sugonarxiv}. Equations (\ref{eq:simplereflectionintro}) and (\ref{eq:refractionintro}) were also used before in the derivation of skew ray tracing equations in optical fibers\cite{sugonSPP}; however, the treatment in this three-page extended conference abstract is limited and does not include the discussion of the numerical aperture. We shall correct these in this paper.

We shall divide the paper into six sections. 
Section \ref{sec:intro} is Introduction. 
In Section \ref{sec:geometricalgebra}, we shall describe the basics of Geometric Algebra and how it can be used for vector rotations in different coordinate systems. 
We shall use this algebra to describe the Laws of Propagation, Reflection, and Refraction in Geometric Optics. 
In Section \ref{sec:skewrays}, we shall derive the distances and angles of propagation of the light rays inside the fiber and compute the fiber's numerical aperture. 
In Section \ref{sec:invariants}, we shall summarize the ray tracing invariants. 
Section \ref{sec:conclusion} is Conclusions.

\section{Geometric Algebra for Geometric Optics} \label{sec:geometricalgebra}

\subsection{Geometric Algebra}

\subsubsection{Vectors and Imaginary Numbers}

Geometric Algebra $\mathcal{C}l_{3,0}$, known as Pauli Algebra,\cite{baylis1} is generated by three spatial unit vectors $\mathbf e_1$, $\mathbf e_2$, and $\mathbf e_3$ which satisfy the orthonormality axioms
\cite{thethreeladies}:
\begin{subequations} 
  \label{eq:garules}
  \begin{align} 
    \label{eq:unitsquare} 
    \mathbf e_j^2 &= 1, 
\\
    \label{eq:flip} 
    \mathbf e_j \mathbf e_k &= -\mathbf e_k \mathbf e_j .
  \end{align}
\end{subequations}
That is, the square of a vector simply is unity and the juxtaposition multiplication product of two distinct unit vectors anti-commute with each other.

We can generalize these multiplication rules for two arbitrary vectors $\mathbf a$ and $\mathbf b$ in 3D using the Pauli Identity\cite{baylis1,lounesto}
\begin{equation}
  \label{eq:pauliidentity}
  \mathbf{a}\mathbf{b} = 
  \mathbf{a} \cdot \mathbf{b} + i(\mathbf{a} \times \mathbf{b}) ,
\end{equation}
where $  i = \mathbf e_1 \mathbf e_2 \mathbf e_3 $.
It can be shown that $i$ is an imaginary number that commutes with both scalars and vectors.

If we express $\mathbf{a}$ as the sum of its perpendicular and parallel vector components with respect to $\mathbf{b}$, then
\begin{equation} 
  \label{eq:aperpparasum}
  \mathbf a = \mathbf{a}_\perp +\mathbf{a}_\parallel .
\end{equation}
Using the Pauli Identity in Eq.~(\ref{eq:pauliidentity}), we can show that
\begin{subequations}
  \label{eq:perpparaflip}
  \begin{align} 
    \label{eq:paraflip}
    \mathbf{a}_\parallel \mathbf{b} &= 
    \mathbf{a}_\parallel \cdot \mathbf{b} = 
    \mathbf{b} \mathbf{a}_\parallel ,
    \\
    \label{eq:perpflip} 
    \mathbf{a}_\perp \mathbf{b} &= 
    \mathbf{a}_\perp \times \mathbf{b} = - \mathbf{b} \mathbf{a}_\perp .
  \end{align}
\end{subequations}
Equation (\ref{eq:perpparaflip}) is the 3D generalization of the 
orthonormality axiom in Eq.~(\ref{eq:garules}).

\subsubsection{Rotations}

Let $\theta$ be a scalar and $\mathbf n$ be a unit vector. Since 
$ (i \mathbf n)^2 = -1 $,
then we may use Euler's Theorem:
\begin{equation} 
  \label{eq:euler}
  e^{i \mathbf{n} \theta} = 
  \cos \theta + i \mathbf n \sin \theta .
\end{equation}

Now, suppose $\mathbf a_\perp$ and  $\mathbf a_\parallel$ are the components of 
$\mathbf a$ perpendicular and parallel to $\mathbf n$. Multiplying these 
from the left of Eq.~(\ref{eq:euler}), and using the commutation and 
anti-commutation rules in Eq.~(\ref{eq:perpparaflip}), we obtain
\begin{subequations} 
  \label{eq:expperparaflip}
  \begin{align} 
    \label{eq:expparaflip} 
    \mathbf{a}_\parallel e^{i \mathbf{n} \theta} &= 
    e^{i \mathbf{n} \theta} \mathbf{a}_\parallel ,
    \\
    \label{eq:expperpflip} 
    \mathbf{a}_\perp e^{i \mathbf{n} \theta} &= 
    e^{-i \mathbf{n} \theta} \mathbf{a}_\perp .
  \end{align}
\end{subequations}
Notice that a change in sign of the exponential's argument occurs only 
when the exponential is multiplied to $\mathbf a_\perp$.

If we expand the left side of Eq.~(\ref{eq:expperpflip}), we get
\begin{equation} 
  \label{eq:aperprotate}
  \mathbf{a}_\perp e^{i \mathbf{n} \theta} = 
  \mathbf{a}_\perp (\cos \theta + i \mathbf n \sin \theta) .
\end{equation}
Distributing the terms, and using the Pauli identity in 
Eq.~(\ref{eq:pauliidentity}), we obtain
\begin{equation} 
  \label{eq:aperpexpand}
  \mathbf{a}_\perp e^{i \mathbf{n} \theta} = 
  \mathbf{a}_\perp \cos \theta - (\mathbf{a}_\perp \times \mathbf n) \sin \theta ,
\end{equation}
since $\mathbf a_\perp \cdot \mathbf n = 0$. Geometrically, 
$\mathbf a_\perp e^{i \mathbf{n} \theta}$ is the vector $\mathbf a_\perp$ 
rotated counterclockwise about a unit vector $\mathbf n$.


\subsubsection{Coordinate Systems}
\label{sec:cylindricalcoordinates}

\begin{figure}
  \centering
   \begin{overpic}[
width=0.8\columnwidth,
tics=5,
trim = 0mm 49.06mm 0mm 49.06mm,
clip=true
page=1
]{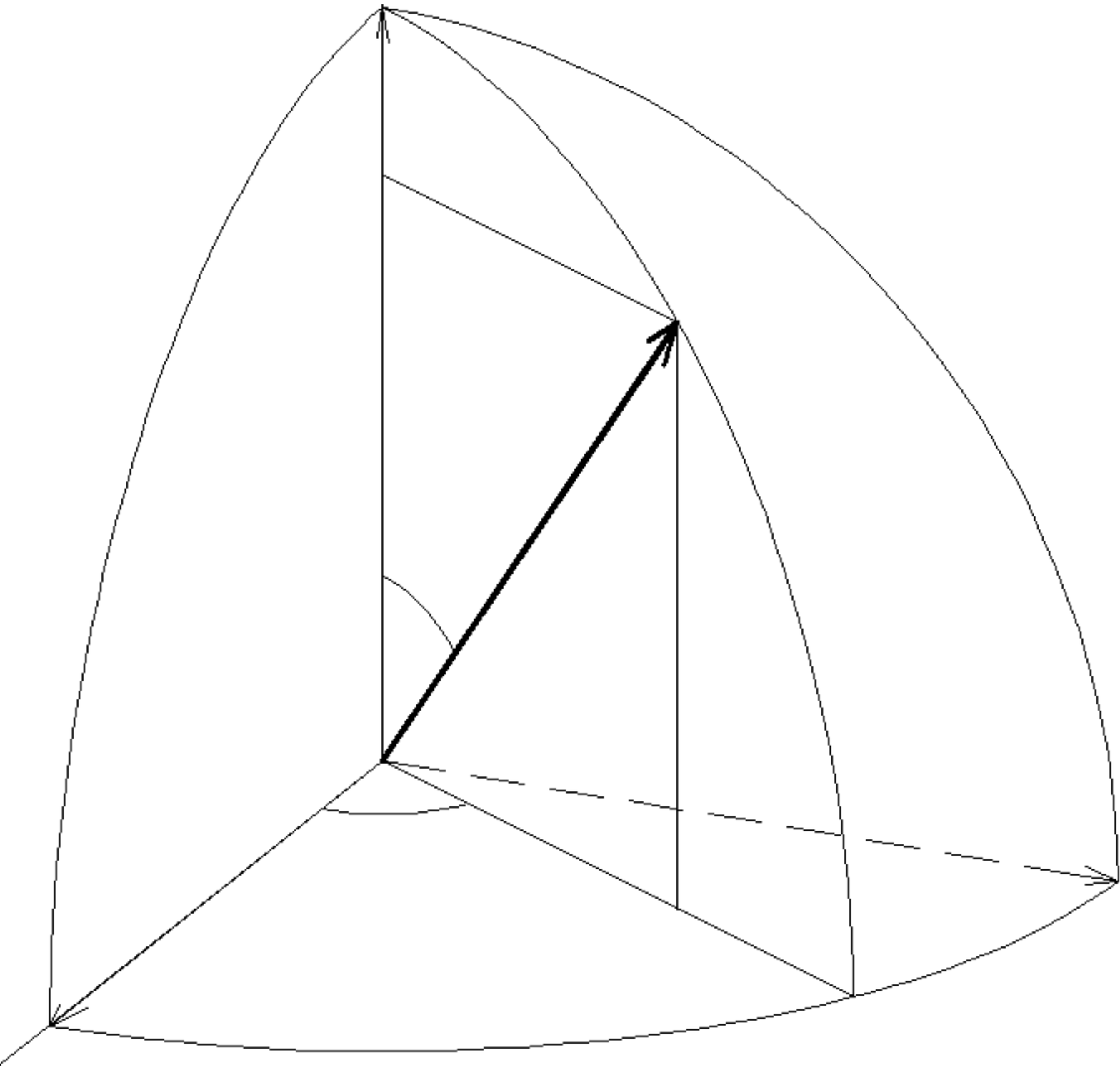}
     \put (61.5,35) {$\cos \theta \, \mathbf e_3$}
     \put (48,17) {{\rotatebox{-31}{$\sin \theta \, \mathbf e_1 e^{i \mathbf e_3 \phi}$}}}
     \put (35,19) {$\phi$}
     \put (37,42) {$\theta$}
     \put (10,5) {$\mathbf e_1$}
     \put (95,20) {$\mathbf e_2$}
     \put (35,88) {$\mathbf e_3$}
  \end{overpic}
  \caption [Cylindrical Coordinate System]{
Relations of the Cylindrical 
    Coordinate System as compared with Cartesian and Spherical Polar. 
}
  \label{fig:cylindrical}
\end{figure}

We claim that the unit radial vector $\mathbf e_r$ in spherical 
coordinates is given by \cite{doranBook}
\begin{equation} 
  \label{eq:unitradial}
  \mathbf e_r = e^{-i \mathbf e_3 \phi/2} \, \mathbf e_3 \, 
e^{i \mathbf e_2 \theta} e^{i \mathbf e_3 \phi/2} .
\end{equation}
To show this, we note that from Eq.~(\ref{eq:aperpexpand}),
\begin{equation} 
  \label{eq:e3rotate}
  \mathbf e_3 e^{i \mathbf e_2 \theta} = 
  \mathbf e_3 \cos \theta +  \mathbf e_1 \sin \theta ,
\end{equation}
so that Eq.~(\ref{eq:unitradial}) may be 
written as 
\begin{equation} 
  \label{eq:unitradialexpand}
  \mathbf e_r = e^{-i \mathbf e_3 \phi/2} \, 
  (\mathbf e_3 \cos \theta +  \mathbf e_1 \sin \theta) \, e^{i \mathbf e_3 \phi/2} .
\end{equation}
Distributing the terms and using the relations in 
Eq.~(\ref{eq:expperparaflip}), we get
\begin{equation} 
  \label{eq:unitradialcylindrical}
  \mathbf e_r = 
  \mathbf e_3 \cos \theta + \mathbf e_1 \sin \theta \, e^{i \mathbf e_3 \phi} .
\end{equation}
Hence,
\begin{equation} 
  \label{eq:radiaialcart}
  \mathbf e_r = \mathbf e_1 \sin \theta \cos \phi + 
  \mathbf e_2 \sin \phi \sin \theta + \mathbf e_3 \cos \theta .
\end{equation}
Equation (\ref{eq:radiaialcart}) is the representation of the unit 
spherical radial vector $\mathbf e_r$ in rectangular coordinates, as shown in Figure \ref{fig:cylindrical}.

\subsection{Geometric Optics} 
\label{sec:geometricoptics}
\subsubsection{Law of Propagation} 
\label{sec:lawofpropagation}

\begin{figure}
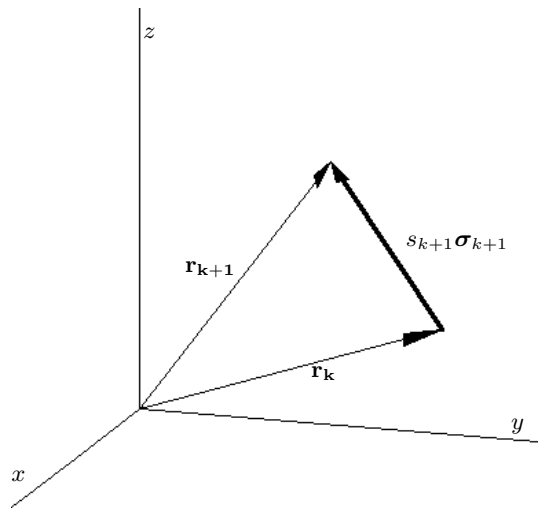

  \centering
   \begin{overpic}[width=.8\columnwidth,tics=5,
clip=true,
trim = 0mm 48.16mm 0mm 48.16mm,
page=2
]{skewrays_graphics}
     \put (33,45) {$\mathbf{r_{k+1}}$}
     \put (57,25) {$\mathbf{r_{k}}$}
     \put (75,50) {$s_{k+1} \bm{\sigma}_{k+1}$}
     \put (0,5) {$x$}
     \put (95,15) {$y$}
     \put (25,90) {$z$}
  \end{overpic}  
  \caption [Propagation Law]{A vector $s_{k+1} \bm{\sigma}_{k+1}$ is defined 
    using two vectors with the same origin point, $\mathbf{r}_k$ and 
    $\mathbf{r}_{k+1}$.}
  \label{fig:proplaw}
\end{figure}

If a light ray moves from its initial position $\mathbf r_k$ at the $k^{th}$ 
interface by a distance $s_{k+1}$ in the direction of the unit vector 
$\bm \sigma_{k+1}$, then the ray's final position $\mathbf r_{k+1}$ at the 
$(k+1)^{th}$ interface is \cite{sugonAJP}
\begin{equation} 
  \label{eq:propagation}
  \mathbf r_{k+1} = 
  \mathbf r_k + s_{k+1} \, \bm \sigma_{k+1} .
\end{equation}
(see Fig.~\ref{fig:proplaw}). Note that we can express $\mathbf r_k$, $\mathbf r_{k+1}$, $\bm \sigma_k$ as the 
sum of $\mathbf e_3$ and $\mathbf e_1 e^{i \mathbf e_3 \phi}$ as given in Eq.~(\ref{eq:unitradialcylindrical}) :
\begin{subequations}
  \label{eq:rkrk1sigform}
  \begin{align}
    \label{eq:rkform}
    \mathbf r_k &= z_k
\mathbf e_3 + 
    r_k \sin \theta_k \mathbf e_1 e^{i \mathbf e_3 \phi_k} , 
    \\
    \label{eq:rk1form}
    \mathbf r_{k+1} &= 
z_{k+1}
\mathbf e_3 
   + 
    r_{k+1} \sin \theta_{k+1} \mathbf e_1 e^{i \mathbf e_3 \phi_{k+1}},  
    \\
    \label{eq:sigmageneral}
    \bm \sigma_{k+1} &= \cos \theta_{\sigma(k+1)} \mathbf e_3 
   + 
    \sin \theta_{\sigma(k+1)} \mathbf e_1 e^{i \mathbf e_3 \phi_{\sigma k+1}} , 
  \end{align}
\end{subequations}
then Eq.~({\ref{eq:propagation}) can be separated into simultaneous equations 
for each component.
\begin{subequations}
  \label{eq:propagationcomponents}
  \begin{align}
    \label{eq:propagationcomponente1}
    r_{k+1} \sin \theta_{k+1} \cos \phi_{k+1} &= r_k  \sin \theta_k \cos \phi_{k} + \notag
    \\& \quad \ 
    s_{k+1} \sin \theta_{\sigma (k+1)} \cos \phi_{\sigma (k+1)} , 
    \\
    \label{eq:propagationcomponente2} 
    r_{k+1} \sin \theta_{k+1} \sin \phi_{k+1} &=     r_k \sin \theta_k \sin \phi_{k} +\notag
    \\& \quad \ 
    s_{k+1} \sin \theta_{\sigma (k+1)} \sin \phi_{\sigma (k+1)} , 
    \\
    \label{eq:propagationcomponente3} 
    r_{k+1} \cos \theta_{k+1} &= 
    r_k \cos \theta_k + s_{k+1} \cos \theta_{\sigma (k+1)} .  
  \end{align}
\end{subequations}

\subsubsection{Law of Reflection} \label{sec:lawofreflection}

Let $\bm \sigma_k$ be the direction of the ray as it hits the $k^{th}$ interface 
with a unit normal vector $\bm \eta_k$. The new direction $\bm \sigma_{k+1}$ 
after it hits the interface is
\cite{doranBook}
  \begin{equation} 
    \label{eq:sigmak1exp}
    \bm \sigma_{k+1} = 
    - e^{-i \bm \eta_k \pi/2} \bm \sigma_k e^{i \bm \eta_k \pi/2} .
\end{equation}
That is, the reflected ray $\bm \sigma_{k+1}$ is the negative of the incident 
ray $\bm \sigma_k$ rotated counterclockwise about $\bm \eta_k$ by an angle 
$\pi$. Expanding the exponentials using the Euler identity Eq.~(\ref{eq:euler}),
then Eq.~(\ref{eq:sigmak1exp}) reduces to
\begin{equation} 
    \label{eq:gareflect}
    \bm \sigma_{k+1} = 
    - \bm \eta_k \bm \sigma_k \bm \eta_k ,
\end{equation}
which is a known form for the reflection law in Geometric Algebra. 
\cite{doranBook}

If we rewrite the product $\bm \eta_k \bm \sigma_k$ in Eq.~(\ref{eq:gareflect}) using the Pauli Identity in Eq.~(\ref{eq:pauliidentity}) and distribute the rightmost $\bm \eta_k$, we get
\begin{equation} 
    \bm \sigma_{k+1} = 
    - (\bm \eta_k \cdot \bm \sigma_k) \bm \eta_k + i (\bm \eta_k \times \bm \sigma_k)\bm \eta_k .
\end{equation}
Applying the Pauli identity again for the second term results to
\begin{equation} 
\label{eq:sigk1expanded}
    \bm \sigma_{k+1} = 
    - \bm \eta_k (\bm \eta_k \cdot \bm \sigma_k) + (\bm \eta_k \times \bm \sigma_k) \times \bm \eta_k ,
\end{equation}
since $i (\bm \eta_k \times \bm \sigma_k) \cdot \bm \eta_k = 0$.
Finally, expanding the triple cross product on right hand side,
Eq.~(\ref{eq:sigk1expanded}) becomes
\begin{equation} 
\label{eq:sigk1dot}
    \bm \sigma_{k+1} = 
    - 2\bm \eta_k (\bm \eta_k \cdot \bm \sigma_k) + \bm \sigma_k .
\end{equation}

Let us define $\beta_k$ as the angle of incidence, so that
\begin{equation}
\bm \eta_k \cdot \bm \sigma_k = c_{\sigma \eta k} \cos \beta_k ,
\end{equation}
where $c_{\sigma \eta k}$ is the concavity function,
\begin{equation}
    \label{eq:concavityfnc}
    c_{\sigma \eta k} = 
    \frac{\bm \sigma_k \cdot \bm \eta_k}{|\bm \sigma_k \cdot \bm \eta_k|} ,
\end{equation}
whose values are $\pm 1$. Thus, Eq.~(\ref{eq:sigk1dot}) reduces to
\begin{equation} 
    \bm \sigma_{k+1} = 
    - 2\bm \eta_k \cos \beta_k + \bm \sigma_k ,
\end{equation}
which is the same expression for the law of reflection in Klein and Furtak. \cite{kleitak,sugonAJP}



\subsubsection{Law of Refraction} 
\label{sec:lawofrefraction}

The Law of Refraction is given by
\begin{equation} 
  \label{eq:snellslaw}
  n \sin \beta = n' \sin \beta' ,
\end{equation}
where $n$ and $n'$ are the indices of refraction at both sides of the interface, while 
$\beta$ and $\beta'$ are the angles of incidence and refraction.

The refraction law in Eq.~(\ref{eq:snellslaw}) may also be reformulated in geometric algebra as 
follows \cite{sugonAJP}:
\begin{equation}
  \label{eq:garefraction}
   \bm \sigma_{k+1} = 
  \bm \sigma_k \, e^{i c_{\sigma \eta} \mathbf e_{\sigma \eta} (\beta- \beta' )} ,
\end{equation}
where $\bm \sigma_k$ is the incident vector, $\bm \sigma_{k+1}$ is the refracted vector, $c_{\sigma \eta k}$ is the concavity function in Eq.~(\ref{eq:concavityfnc}), and $\mathbf e_{\sigma \eta k}$ is the axis of rotation
\begin{equation}
    \label{eq:axisofrotationfnc}
    \mathbf e_{\sigma \eta k} = 
    \frac{\bm \sigma_k \times \bm \eta_k}{|\bm \sigma_k \times \bm \eta_k|} ,
\end{equation}
with $\bm \eta_k$ is the unit vector normal to the surface. Geometrically, Eq.~(\ref{eq:garefraction}) says that the refracted ray is the incident ray rotated either clockwise (if the concavity function 
is negative/interface is convex) or counterclockwise (if the concavity function is 
positive/interface is concave) about $\mathbf e_{\sigma \eta k}$ by the difference of the 
incident angle and the refracted angle, $\beta - \beta'$. 
We will use primed variables for the refracted values.
Also, we shall mostly deal with concave surfaces inside the fiber, so that we can immediately 
set $c_{\sigma \eta k} = +1$.


\section{Skew Rays in Optical Fibers} 
\label{sec:skewrays}

%
%

\begin{figure}
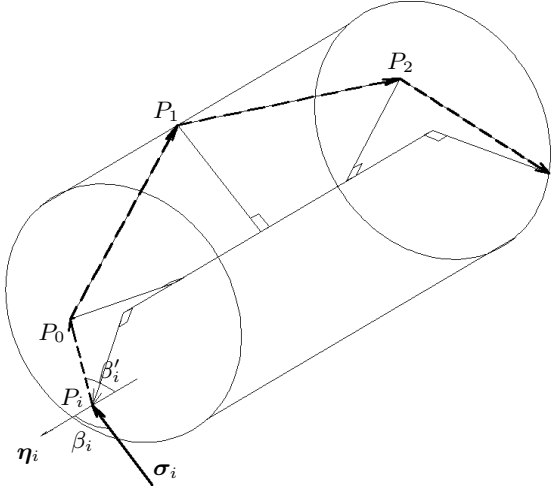
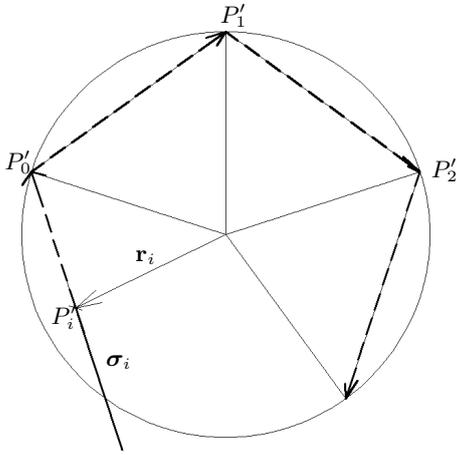

    \centering
    \begin{subfigure}[b]{.5\textwidth}
        \centering
   \begin{overpic}[width=.8\columnwidth,tics=5,clip=true,trim=0mm 54.55mm 0mm 54.55mm,page=4]{skewrays_graphics}
     \put (17,20) {$\beta_i'$}
     \put (12,7) {$\beta_i$}
     \put (2,5) {$\bm \eta_i$}
     \put (27,2) {$\bm  \sigma_i$}
     \put (10,15) {$P_i$}
     \put (6,27) {$P_0$}
     \put (27,68) {$P_1$}
     \put (70,77) {$P_2$}
  \end{overpic}  
     \caption [Divisions of the Paper]{ 
The incident ray $\bm \sigma_i$ strikes the point $P_i$ entrance end of the fiber where the normal vector is $\bm \eta_i$. The angle of incidence is $\beta_i$ and the angle of refraction is $\beta_i'$. From point $P_i$, the light ray enters the fiber and strikes the cylindrical walls at points $P_0$, $P_1$, and $P_2$.}
   \label{fig:main}
    \end{subfigure}
    \begin{subfigure}[b]{.5\textwidth}
        \centering
\vspace{2em}
   \begin{overpic}[width=0.6\columnwidth,tics=5, clip=true,trim=0mm 39.98mm 0mm 39.98mm,page=5
]{skewrays_graphics}
		\put (27,45) {$\mathbf r_i$}
		\put (20,20) {$\bm  \sigma_i$}
     \put (7,30) {$P_i'$}
     \put (-4,67) {$P_0'$}
     \put (47,102) {$P_1'$}
     \put (97,65) {$P_2'$}
  \end{overpic}  
   \caption [Divisions of the Paper]{Top view of Fig.~(\ref{fig:main}), with the points $P_0'$, $P_1'$, and $P_2'$ as the projections of the points $P_0$, $P_1$, and $P_2$ on the fiber's entrance.}
   \label{fig:main2}
    \end{subfigure}
  \caption [Divisions of the Paper]{The propagation of the light ray inside the optical fiber in perspective view and top view}
     \label{fig:main3}
\end{figure}

\subsection{Refraction at Point $P_i$} 
\label{sec:refractionatA}


Let $\bm  \sigma_i$ be the direction of the initial ray as it strikes the flat end of the fiber:
\begin{equation} 
  \label{eq:sigmaneg1defn}
  \bm  \sigma_i = 
  \sin  \theta_{ \sigma_i } \, \mathbf e_1 e^{ i \mathbf e_3 \phi_{ \sigma_i } } + 
  \cos \theta_{ \sigma_i } \, \mathbf e_3 ,
\end{equation}
which is similar in form to Eq.~(\ref{eq:sigmageneral}). If the ray $\bm  \sigma_i$ strikes 
the fiber at the position $\mathbf  r_i$, then the outward normal vector at the surface is
\begin{equation} 
  \label{eq:eta0}
  \bm  \eta_i = - \mathbf e_3 .
\end{equation}

Multiplying Eq.~(\ref{eq:sigmaneg1defn}) and Eq.~(\ref{eq:eta0}), and using the exponential 
identities in Eq.~(\ref{eq:perpparaflip}), we get
\begin{equation} 
  \label{eq:signeg1eta0}
  \bm  \sigma_i \bm  \eta_i 
  =  \sin \theta_{ \sigma_i } \, i \mathbf e_2 e^{ i \mathbf e_3 \phi_{ \sigma_i } } - 
  \cos \theta_{ \sigma_i } .
\end{equation}
Expanding the left-hand side of Eq.~(\ref{eq:signeg1eta0}) using the Pauli Identity in Eq.~(\ref{eq:pauliidentity}), and separating the scalar and vector parts, we obtain
\begin{subequations} 
  \label{eq:sigmaneg1pauli}
  \begin{align} 
    \label{eq:sigmaneg1dot} 
    \bm  \sigma_i \cdot \bm  \eta_i &= 
    -\cos \theta_{ \sigma_i } ,
    \\
    \label{eq:sigmaneg1cross} 
    \bm  \sigma_i \times \bm  \eta_i &= 
    \sin \theta_{ \sigma_i } \, \mathbf e_2 e^{ i \mathbf e_3 \phi_{ \sigma_i } } .
  \end{align}
\end{subequations}
Substituting these into the definition of the the concavity function  and the rotation 
axis in Eqs.~(\ref{eq:concavityfnc}) and (\ref{eq:axisofrotationfnc}), we arrive at
\begin{subequations}
  \label{eq:refractionparameteranser}
  \begin{align}
    \label{eq:concavityans}
    c_{\sigma \eta i} &= 
    - \frac{\cos \theta_{ \sigma_i }}{|\cos \theta_{ \sigma_i }|} = - 1 ,
    \\
    \label{eq:axisofrotationans}
    \mathbf e_{\sigma \eta i} &= 
    \mathbf e_2 e^{i \mathbf e_3 \phi_{ \sigma_i }} ,
  \end{align}
\end{subequations}
since $\theta_{ \sigma_i } > 0$.

We can find the angle $\beta_i$ between the unit vectors $\bm  \sigma_i$ and 
$\bm  \eta_i$ using their dot product:
\begin{equation} 
  \label{eq:sigmaetadotsimp}
  \bm  \sigma_i \cdot \bm  \eta_i = 
  \cos \beta_i = \cos \theta_{ \sigma_i } ,
\end{equation}
so that
\begin{equation} 
  \label{eq:beta1}
  \beta_i = \theta_{ \sigma_i } .
\end{equation}
Using Snell's Law, the refracted angle $\beta_i'$ is given by
\begin{equation}
  \label{eq:beta1snells}
  \beta_i' = 
  \sin^{-1} \left (\frac{n}{n'} \sin \beta_i \right) ,
\end{equation}
where we set $n=1$ for air (see Fig.~\ref{fig:main3}).

The law of refraction as given by Eq.~(\ref{eq:garefraction}) for $k=0$ is
\begin{equation} 
  \label{eq:sig0refracted}
  \bm \sigma_0 = 
  \bm  \sigma_i e^{i (\beta_i - \beta_i' ) \mathbf e_{\sigma \eta i}} ,
\end{equation}
where $c_{\sigma \eta 1} = 1$. 
We can express $\bm  \sigma_i$ as a rotation of $\mathbf e_3$ about the 
axis $\mathbf e_{\sigma \eta 1}$ by a clockwise angle $\beta_i$,
\begin{equation}
  \label{eq:signeg1exp}
  \bm  \sigma_i = 
  \mathbf e_3 e^{- i \mathbf e_{\sigma \eta i} \beta_i} ,
\end{equation}
so that Eq.~(\ref{eq:sig0refracted}) reduces to
\begin{equation} 
  \label{eq:sig0resultexp}
  \bm \sigma_0 = 
  \mathbf e_3 e^{- i \beta_i' \mathbf e_{\sigma \eta i}} .
\end{equation}

Expanding the exponential in Eq.~(\ref{eq:sig0resultexp}), we get
\begin{equation} 
  \label{eq:sig0soln}
  \bm \sigma_0 = 
  \sin \beta_i' \, \mathbf e_1 e^{ i \mathbf e_3 \phi_{ \sigma_i } } + 
  \cos \beta_i' \, \mathbf e_3 .
\end{equation}
On the other hand, using the expansion of $\bm \sigma_0$ in 
Eq.~(\ref{eq:sigmageneral}), we have
\begin{equation} 
  \label{eq:sig0generalform}
  \bm \sigma_{0} = 
  \cos \theta_{ \sigma_0 } \mathbf e_3 + 
  \sin \theta_{ \sigma_0 } \mathbf e_1 e^{i \mathbf e_3 \phi_{ \sigma_0 }} .
\end{equation}
Thus, equating Eqs.~(\ref{eq:sig0soln}) and (\ref{eq:sig0generalform}), we get
\begin{subequations} 
  \label{eq:anglesigmarelations}
  \begin{align} 
    \label{eq:thetasigma0} 
    \theta_{ \sigma_0 } &= \beta_i' ,
    \\
    \label{eq:phisigma0} 
    \phi_{ \sigma_0 } &= \phi_{ \sigma_i } .
  \end{align}
\end{subequations}
which is the same relations in \cite{sugonSPP}, once we have changed the subscripts 
to match their notation.


\subsection{Propagation from $P_i$ to $P_0$} 
\label{sec:propagationAtoB}

\begin{figure}
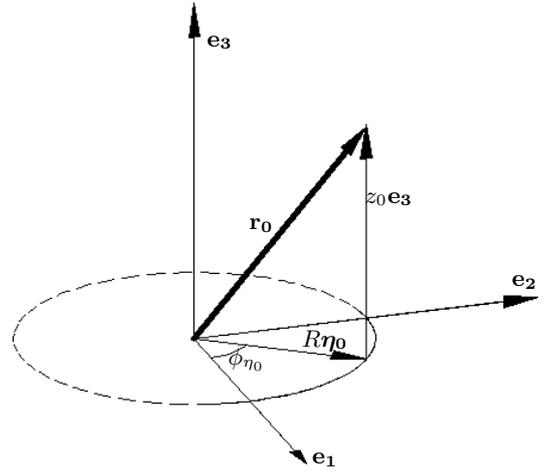

   \centering
   \begin{overpic}[
width=.8\columnwidth,tics=5,
trim=0mm 55.91mm 0mm 55.91mm,
clip=true,
page=3
]{skewrays_graphics}
     \put (45,45) {$\mathbf{r_{0}}$}
     \put (67,50) {$z_0 \mathbf{e_3}$}
     \put (55,22.5) {$R \bm{\eta_{0}}$}
     \put (41,19) {$\phi_{ \eta_0 }$}
     \put (57,0) {$\mathbf{e_1}$}
     \put (95,34) {$\mathbf{e_2}$}
     \put (37,80) {$\mathbf{e_3}$}
  \end{overpic}  
   \caption [Illustration of vector r1]{Vector $\mathbf  r_0$ can be expressed as the sum of a rotating vector on the $xy$ plane, $R\, \bm  \eta_0$, and a vertical component, $z_0 \mathbf e_3$.}
   \label{fig:r1_illus}
\end{figure}

We define the entry point of the light ray when it intersects the base of the 
cylinder as $\mathbf  r_i$:
\begin{equation} 
  \label{eq:r0defn}
  \mathbf  r_i = 
  \rho_i \, \mathbf e_1 e^{ i \mathbf e_3 \phi_i } = 
  \rho_i \cos \phi_i \mathbf e_1 + \rho_i \sin \phi_i \mathbf e_2 .
\end{equation}
From position $\mathbf  r_i$, the ray moves in the direction of $\bm \sigma_0$ 
defined in Eq.~(\ref{eq:sig0generalform}), until it strikes the walls of the cylinder 
at position $\mathbf  r_0$, so that
\begin{equation} 
  \label{eq:r1s0defn}
  \mathbf  r_0 = s_0 \bm \sigma_0 + \mathbf  r_i ,
\end{equation}
where $s_0$ is the distance of propagation. If we define the radius of the cylinder 
to be $R$, then the position $\mathbf  r_0$ can be written as
\begin{equation} 
  \label{eq:r1defn}
  \mathbf  r_0 = 
  R\, \mathbf e_1 e^{ i \mathbf e_3 \phi_{ \eta_0 } } + z_0 \mathbf e_3 ,
\end{equation}
where $\phi_{ \eta_0 }$ is the azimuthal angle and $z_0$ is the distance along the 
axis, as illustrated in Figure \ref{fig:r1_illus}. Substituting Eq.~(\ref{eq:r1s0defn}) into Eq.~(\ref{eq:r1defn}), we get
\begin{equation}
  \label{eq:threeunknowns}
  s_0 \bm \sigma_0 + \mathbf  r_i = 
  R\, \mathbf e_1 e^{ i \mathbf e_3 \phi_{ \eta_0 } } + z_0 \mathbf e_3 .
\end{equation}
Notice that there are three unknowns in this equation: $s_0$, $\phi_{ \eta_0 }$, and $z_0$.

Let us separate the axial and radial components of Eq.~(\ref{eq:threeunknowns}) to obtain
\begin{subequations}
  \label{eq:axialradialthreeunkn}
  \begin{align}
    \label{eq:axialthreeunkn}
    s_0 \cos \theta_{ \sigma_0 } \mathbf e_3 &= 
    z_0 \mathbf e_3 ,
    \\
    \label{eq:radialthreeunkn}
    s_0 \, \sin \theta_{ \sigma_0 } \mathbf e_1 e^{i \mathbf e_3 \phi_{ \sigma_0 }} +
    \rho_i \, \mathbf e_1 e^{ i \mathbf e_3 \phi_i } &= 
    R\, \mathbf e_1 e^{ i \mathbf e_3 \phi_{ \eta_0 } } .
  \end{align}
\end{subequations}
Equation~(\ref{eq:axialthreeunkn}) gives us our first equation relating $s_0$ and $z_0$:
\begin{equation}
  \label{eq:s0z1relation}
 z_0 =  s_0 \cos \theta_{ \sigma_0 }.
\end{equation}

Now, squaring both sides of Eq.~(\ref{eq:radialthreeunkn}) and using the exponential identities in Eq.~(\ref{eq:expperparaflip}), we get
\begin{equation}
  s_0^2 \, \sin^2 \theta_{ \sigma_0 } + \rho_i^2 + 
  2 s_0 \rho_i \, \sin \theta_{ \sigma_0 } \, \cos (\phi_i  - \phi_{ \sigma_0 })
  = 
  R^2 .
\end{equation}
Solving for $s_0 \sin \theta_{ \sigma_0 }$ using the quadratic formula results to
\begin{align}
\label{eq:s0plmin}
  s_0\, \sin \theta_{ \sigma_0 } &= 
  \pm \sqrt{ R^2 - \rho_i^2 \sin^2(\phi_i - \phi_{ \sigma_0 }) } \notag
\\
 & \qquad - 
  \rho_i\,\cos(\phi_i - \phi_{ \sigma_0 }) ,
\end{align}
after simplifying the terms.

Notice that there are two possible solutions:
\begin{subequations}
  \begin{align}
  \label{eq:s0plus}
  s_{0\pm} &= \frac{1}{\sin \theta_{ \sigma_0 } } 
\Big [ \pm
  \sqrt{ R^2 - \rho_i^2 \sin^2(\phi_i - \phi_{ \sigma_0 }) }  \notag
\\ & \qquad 
- 
  \rho_i\,\cos(\phi_i - \phi_{ \sigma_0 }) 
  \Big ] , 
  \end{align}
\end{subequations}
The distance $s_{0+}$ is the distance traveled by the light ray inside the fiber from the fiber's entrance to the cylindrical interface, while the distance $s_{0-}$ is the distance traveled by the light ray if it propagates opposite to the direction $\bm \sigma_0$. The difference between these two distances is 
\begin{equation}
  s_{0+} - s_{0-} = 
  \frac{2}{\sin \theta_{ \sigma_0 } } 
  \sqrt{ R^2 - \rho_i^2 \sin^2(\phi_i - \phi_{ \sigma_0 }) } ,
\end{equation}
which is a new result (see Fig. \ref{fig:s0pls0min}). 
If $\rho_i = 0$, then we get the meridional case:
\begin{equation}
  s_{0+} - s_{0-} = 
  \frac{2R}{\sin \theta_{ \sigma_0 } } .
\end{equation}
On the other hand, if $\rho_i = R$, 
\begin{equation}
  s_{0+} - s_{0-} = 
  \frac{2 R | \cos(\phi_i - \phi_{ \sigma_0 }) | }{\sin \theta_{ \sigma_0 } }  ,
\end{equation}
which is similar to the form of the wall-to-wall propagation distance given by Cozannet and Treheux\cite{Cozannet}, except that for our case, $s_{0-}$ is a virtual backward propagation from the fiber entrance.

\begin{figure}
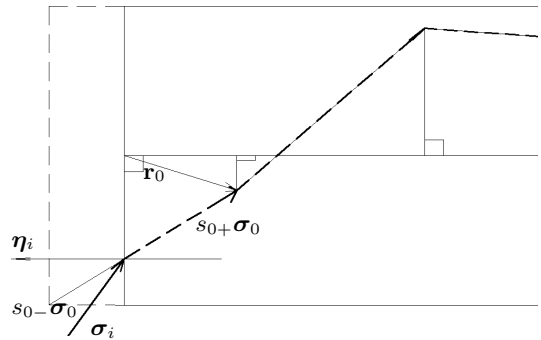

   \centering
  \begin{overpic}[width=.8\columnwidth,tics=5,  clip=true,trim=0mm 82.34mm 0mm 82.34mm,page=6]{skewrays_graphics}
     \put (15,0) {$\bm  \sigma_i$}
     \put (0,17) {$\bm \eta_i$}
     \put (25,30) {$\mathbf  r_0$}
     \put (35,20) {$s_{0+}\bm \sigma_{0}$}
     \put (0,4) {$s_{0-}\bm \sigma_{0}$}
  \end{overpic}
   \caption {If we extend $s_{0+}\bm \sigma_{0}$ backwards, we obtain a vector of length $s_{0+}$, which will terminate at the fiber's wall, extended in the $-\mathbf e_3$ direction.}
   \label{fig:s0pls0min}
\end{figure}

Lastly, we solve for $\phi_{ \eta_0 }$. Separating Eq.~(\ref{eq:radialthreeunkn}) into the components of $\mathbf e_1$ and $\mathbf e_2$, we get
\begin{subequations}
\label{eq:phieta1componentsa}
  \begin{align}
\label{eq:phieta1componentsa}
    s_0\, \sin  \theta_{ \sigma_0 } \, \cos  \phi_{ \sigma_0 }  + 
    \rho_i\, \cos  \phi_i  &= 
    R \, \cos  \phi_{ \eta_0 }  ,
    \\
\label{eq:phieta1componentsb}
    s_0\,\sin  \theta_{ \sigma_0 } \, \sin  \phi_{ \sigma_0 }  + 
    \rho_i\, \cos  \phi_i  &= 
    R \, \sin  \phi_{ \eta_0 }  .
  \end{align}
\end{subequations}
Dividing Eq.~(\ref{eq:phieta1componentsa}) by Eq.~(\ref{eq:phieta1componentsb}), we obtain
	\begin{equation} \label{eq:phieta1solution}
		\tan \phi_{ \eta_0 } = \dfrac{ y_{ \eta_0 } }{ x_{ \eta_0 } } ,
	\end{equation}
where
\begin{subequations}
\label{eq:phietaxy}
\begin{align}
	x_{ \eta_0 } &= s_0\, \sin\, \theta_{ \sigma_0 } \, \cos \phi_{ \sigma_0 }  + \rho_i\, \cos  \phi_i  ,
	\\
   y_{ \eta_0 } &= s_0\, \sin \theta_{ \sigma_0 } \, \sin  \phi_{ \sigma_0 }  + \rho_i\, \sin  \phi_i  .
\end{align}
\end{subequations}
We note that $s_0 \sin \theta_{\sigma_0}$ is given in Eq.~(\ref{eq:s0plmin}), so that the rectangular coordinates 
$x_{\eta_0}$ and $y_{\eta_0}$
of the normal vector $\bm \eta_0$
are already defined in terms of the initial parameters.
Thus,
the cylindrical coordinates of the normal vector $\bm \eta_0$ can now be determined, since the radius $R$ of the fiber is constant and the azimuthal angle $\phi_{\eta_0}$ is given by Eq.~(\ref{eq:phieta1solution}).

	 \subsection{Refraction at $P_0$: Numerical Aperture}
	 
At point $P_0$, on the fiber's cylindrical wall, the incident ray is $\bm \sigma_0$ defined in Eq.~(\ref{eq:sig0generalform}), and the normal vector is 
\begin{equation}
\bm \eta_0 = \mathbf e_1 e^{i \mathbf e_3 \phi_{\eta_0}} .
\end{equation}
Their product is
\begin{equation}
	\bm \sigma_0 \bm  \eta_0 = \sin \theta_{ \sigma_0 } e^{i \mathbf e_3 (\phi_{ \sigma_0 } - \phi_{ \eta_0 })} - i \cos \theta_{ \sigma_0 } \mathbf e_2 e^{i \mathbf e_3 \phi_{ \eta_0 }} ,
\end{equation}
where we used the exponential indentities in Eq.~(\ref{eq:expperparaflip}).
Separating the scalar and imaginary vector parts, we get
\begin{align}
\label{eq:sig0eta0dot}
	\bm \sigma_0 \cdot \bm \eta_0 &= \sin \theta_{ \sigma_0 } \cos (\phi_{ \sigma_0 } - \phi_{ \eta_0 }) ,
\\
	\bm \sigma_0 \times \bm \eta_0 &= 
	\mathbf e_1 \cos \theta_{ \sigma_0 } \sin \phi_{ \eta_0 } - 
  \mathbf e_2 \cos \theta_{ \sigma_0 }  \cos \phi_{ \eta_0 }  \notag
\\ & \quad	 \,\,
  + \mathbf e_3 \sin \theta_{ \sigma_0 } \sin (\phi_{ \sigma_0 } - \phi_{ \eta_0 }) , 
\end{align}
after removing the imaginary number $i$ in the second equation. If we define $\psi'$ as the angle between $\bm \sigma_0$ and $\bm \eta_0$, then Eq.~(\ref{eq:sig0eta0dot}) yields
\begin{equation}
\label{eq:cospsipri}
\cos \psi' = \sin \theta_{ \sigma_0 } \cos (\phi_{ \sigma_0 } - \phi_{ \eta_0 }) ,
\end{equation}
which is essentially the same as that of Potter\cite{Potter} and Senior\cite{senior}.

Since $\psi'$ is also the angle of incidence at point $P_0$, then by Snell's Law we have
\begin{equation}
n' \sin \psi' = n \sin \psi ,
\end{equation}
where $\psi$ is the angle of refraction outside at point~$P_0$.
For the ray to be trapped, we set $\psi = \pi / 2$, so that
\begin{equation}
\label{eq:sinpsipri}
\sin \psi' = \dfrac{n}{n'} .
\end{equation}
Combining Eq.~(\ref{eq:cospsipri}) and Eq.~(\ref{eq:sinpsipri}),
\begin{equation}
1 = \dfrac{n^2}{n'^2} + \sin^2 \theta_{ \sigma_0 } \cos^2 (\phi_{ \sigma_0 } - \phi_{ \eta_0 }) ,
\end{equation}
then solving for $\sin \theta_{ \sigma_0 }$, we get
\begin{equation}
\label{eq:sinthetasig0}
\sin \theta_{ \sigma_0 } = \dfrac{\displaystyle \sqrt{n'^2 - n^2}}{n' | \cos (\phi_{ \sigma_0 } - \phi_{ \eta_0 }) | } .
\end{equation}

Now, the numerical aperture NA of an optical fiber is defined as
\begin{equation}
\label{eq:numaperture}
\mathrm{NA} = n' \sin \theta_{ \sigma_0 } .
\end{equation}
We combine Eqs.~(\ref{eq:sinthetasig0}) and (\ref{eq:numaperture}) to get
\begin{equation}
\label{eq:numaptfull}
\mathrm{NA} = \dfrac{\displaystyle \sqrt{n'^2 - n^2}}{|\cos (\phi_{ \sigma_0 } - \phi_{ \eta_0 })|} .
\end{equation}
Equation (\ref{eq:numaptfull}) is the expression for the numerical aperture for skew rays in optical fibers as given by Potter\cite{Potter,Potter2} and Senior\cite{senior}.

For the meridional approximation, we set $\phi_{\eta_0} = \phi_{\sigma_0}$, so that Eq.~(\ref{eq:numaptfull}) reduces to the known standard form for the numerical aperture as given by Klein and Furtak\cite{kleitak} and Potter\cite{Potter}:
\begin{equation}
\label{eq:meridionalnumapt}
\mathrm{NA} = \sqrt{n'^2 - n^2} .
\end{equation} 

If we expand the denominator in Eq.~(\ref{eq:numaptfull}), we get
\begin{equation}
\cos (\phi_{ \sigma_0 } - \phi_{ \eta_0 }) = \cos \phi_{ \sigma_0 } \cos \phi_{ \eta_0 } + \sin \phi_{ \sigma_0 } \sin \phi_{ \eta_0 } .
\end{equation}
Since
\begin{subequations}
\begin{align}
\cos \phi_{\eta_0} &= {x_{\eta_0} /R} ,
\\
\sin \phi_{\eta_0} &= {y_{\eta_0} /R},
\end{align}
\end{subequations}
then using Eq.~(\ref{eq:phietaxy}) together with trigonometric indentities, we obtain
\begin{equation}
\label{eq:numaptdenom}
\cos (\phi_{ \sigma_0 } - \phi_{ \eta_0 }) = 
\dfrac{s_0}{R}\, \sin\, \theta_{ \sigma_0 } + \dfrac{\rho_i}{R} \cos (\phi_i - \phi_{ \sigma_0 }) .
\end{equation}

Substituting Eq.~(\ref{eq:numaptdenom}) back into Eq.~(\ref{eq:numaptfull}) results to
\begin{equation}
\mathrm{NA} =  \dfrac{\displaystyle \sqrt{n'^2 - n^2}}{\left |
\dfrac{s_0}{R}\, \sin\, \theta_{ \sigma_0 } + \dfrac{\rho_i}{R} \cos (\phi_i - \phi_{ \sigma_0 })
\right |} .
\end{equation}
Combining this with the expression for $s_0$ in Eq.~(\ref{eq:s0plus})
and using the identity $\phi_{ \sigma_0 } = \phi_{ \sigma_i }$ from Eq.~(\ref{eq:phisigma0}),
we arrive at
\begin{equation}
\label{eq:numaptgamma}
\mathrm{NA} = 
\chi \sqrt{n'^2 - n^2} ,
\end{equation}
where
\begin{equation}
\label{eq:chi}
\chi = \dfrac{1}{
{\sqrt{ 1 - \dfrac{\rho_i^2}{R^2} \sin^2 (\phi_i - \phi_{ \sigma_i }) }
}} .
\end{equation}
Equation (\ref{eq:numaptgamma}) forms the expression for the numerical aperture NA for the skew rays in optical fibers. 
Here we introduce the skew factor $\chi$ which is defined in terms of the initial ray parameters at the fiber entrance: the position $\mathbf r_i = (\rho_i, \phi_i, 0)$ in cylindrical coordinates where the incident ray $\bm \sigma_i = (1, \phi_{ \sigma_i }, \theta_{\sigma_i})$ in spherical coordinates. Notice that the skew factor depends on the azimuthal angle $\phi_{ \sigma_i }$ of the incident ray, but not on the polar angle $\theta_{\sigma_i}$.

The expression for the numerical aperture NA in terms of the skew factor $\chi$ is a new result.
To arrive at the meridional case in Eq.~(\ref{eq:meridionalnumapt}), we set the radial distance $\rho_i = 0$, to get the skew factor $\chi = 1$. 
On the other hand, if we set the initial position at the edge of the fiber, i.e. $\rho_i = R$, we get 
\begin{equation}
\chi=
\dfrac{1}{\sqrt{\cos^2 (\phi_{ \sigma_0 } - \phi_{ \eta_0 })}} ,
\end{equation}
which leads to Eq.~(\ref{eq:numaptfull}), the numerical aperture for skew rays given in the literature. Fig. \ref{fig:numapt} shows the plot of the skew factor $\chi$ as a function of the normalized radial distance $\rho_i/R$ and the difference of the azimuthal angles  $\phi_i - \phi_{ \sigma_i }$.



\begin{figure}
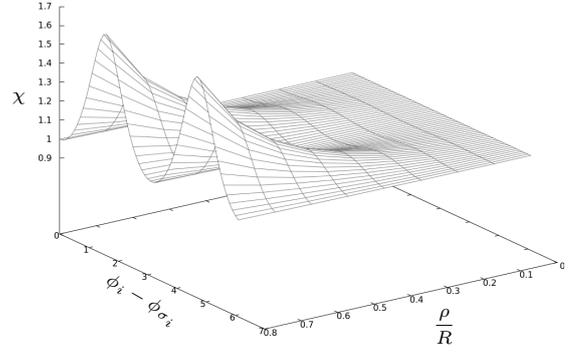

  \centering
   \begin{overpic}[width=.8\columnwidth,tics=5,
trim = 0mm 81.69mm 0mm 81.69mm,
clip=true,
page=8
]{skewrays_graphics}
     \put (75,1) {\scriptsize $\dfrac{\rho}{R}$}
     \put (12, 10) {\rotatebox{-30}{\scriptsize $\phi_i - \phi_{ \sigma_i }$}}
     \put (-5,45) {\scriptsize $\chi$}
  \end{overpic}
  \caption [Numerical Aperture]{The skew factor $\chi$ as a function of the ratios of the relative radii $\rho/R$ and the azimuthal angular difference $\phi_i - \phi_{ \sigma_i }$ }
    \label{fig:numapt}
\end{figure}

\subsection{Reflection at $P_0$} 
\label{sec:reflectionatB}

At position $\mathbf  r_0$ on the interface, the light ray's propagation vector is $\bm \sigma_0$ and the normal vector to the interface is $\bm  \eta_0$: 
\begin{subequations}
  \begin{align}
    \label{eq:sig0defn}
    \bm \sigma_{0} &= 
    \cos \theta_{ \sigma_0 } \mathbf e_3 + 
    \sin \theta_{ \sigma_0 } \mathbf e_1 e^{i \mathbf e_3 \phi_{ \sigma_0 }} ,
    \\
    \label{eq:eta1defn}
    \bm  \eta_0 &= \mathbf e_1 e^{i \mathbf e_3 \phi_{ \eta_0 } } .
  \end{align}
\end{subequations}
To obtain the new propagation vector $\bm \sigma_1$ after reflection, we use the law of reflection
in Eq.~(\ref{eq:gareflect}):
\begin{equation} 
  \label{eq:sig1reflect}
  \bm \sigma_1 = 
  - \bm  \eta_0 \bm \sigma_0 \bm  \eta_0 .
\end{equation}
Substituting the expressions for $\bm  \eta_0$ in Eq.~(\ref{eq:eta1defn}), 
and $\bm \sigma_0$ in Eq.~(\ref{eq:sig0defn}) into Eq.~(\ref{eq:sig1reflect}), 
yields
\begin{equation} 
  \label{eq:sig1reflected}
  \bm \sigma_1  = 
  - \sin \theta_{ \sigma_0 } \, \mathbf e_1 
  e^{ i \mathbf e_3 (2\phi_{ \eta_0 } - \phi_{ \sigma_0 } ) } + 
  \cos \theta_{ \sigma_0 } \, \mathbf e_3 ,
\end{equation}
after distributing the terms and simplifying the result. Eq.~(\ref{eq:sig1reflected}) is the desired equation for the new propagation vector $\bm \sigma_1$.

Using the definition of $\bm \sigma_1$ as written in Eq.~(\ref{eq:sigmageneral}),
\begin{equation} 
  \label{eq:sig1defn}
  \bm \sigma_1 = 
  \sin \theta_{ \sigma_1 } \, \mathbf e_1 e^{ i \mathbf e_3 \phi_{ \sigma_1 } } 
  + \cos \theta_{ \sigma_1 } \, \mathbf e_3 ,
\end{equation}
and comparing this with the expression for $\bm \sigma_1$ in 
Eq.~(\ref{eq:sig1defn}), we obtain \cite{sugonSPP}
\begin{subequations} 
  \label{eq:anglerelations}
  \begin{align}
    \label{eq:thetarelation} 
    \theta_{ \sigma_1 } &= - \theta_{ \sigma_0 } ,
    \\
    \label{eq:phirelation} 
    \phi_{ \sigma_1 } &= 2\phi_{ \eta_0 } - \phi_{\sigma0} .
    \end{align}
\end{subequations}
Notice that Eq.~(\ref{eq:thetarelation}) is not proper as we 
previously defined $\theta$ to have values between $0$ to $\pi$.

In order to avoid having to write the negative sign before the $\theta$, 
we can use the following relation derived from the Euler identity:
	\begin{equation} \label{eq:eulerpi}
		e^{i \mathbf e_3 (\pm \pi)} = \cos (\pm \pi) + i \mathbf e_3 \sin (\pm \pi) = -1 ,
	\end{equation}
so that Eq.~(\ref{eq:sig1reflected}) becomes
\begin{equation}
  \bm \sigma_1  = 
  \sin \theta_{ \sigma_0 } \, 
  \mathbf e_1 e^{ i \mathbf e_3 (2\phi_{ \eta_0 } - \phi_{\sigma0} \pm \pi ) } 
  + \cos \theta_{ \sigma_0 } \, \mathbf e_3 .
\end{equation}
Hence,
\begin{subequations} 
  \label{eq:newanglerelations}
  \begin{align} 
    \label{eq:newthetarelation} \theta_{ \sigma_1 } &= \theta_{ \sigma_0 } ,
    \\
\label{eq:newphisigmarelation}
\phi_{ \sigma_1 } &= 2\phi_{ \eta_0 } - \phi_{\sigma0} \pm \pi ,
  \end{align}
\end{subequations}
which is similar to what we have derived in a previous paper\cite{sugonSPP}.
Notice that Eq.~(\ref{eq:newthetarelation}) is now in the proper form of $\theta$, except that the sign of $\pi$ in Eq.~(\ref{eq:newphisigmarelation}) must be determined for individual cases.

\subsection{Propagation from $P_0$ to $P_1$}
\label{sec:propagationatC}

From the propagation law in Eq.~(\ref{eq:propagation}), we know that the final position of the ray depends on its initial position $\mathbf  r_0$:
\begin{equation} 
  \label{eq:r2prop}
  \mathbf  r_1 = s_1 \bm \sigma_1 + \mathbf  r_0 ,
\end{equation}
where $s_1$ and $\bm \sigma_1$ is the ray's propagating distance and direction, respectively.
This is shown in Figure \ref{fig:secondprop}.
If we use the expression of $\mathbf  r_0$  in Eq.~(\ref{eq:r1defn}) and express
$\mathbf  r_1$ in a similar form, then Eq.~(\ref{eq:r2prop}) becomes
\begin{equation} \label{eq:threeunknownsC}
  s_1 \bm \sigma_1 + R \mathbf e_1 e^{i \mathbf e_3 \phi_{ \eta_0 }} + z_0 \mathbf e_3 = 
  R \mathbf e_1 e^{i \mathbf e_3 \phi_{ \eta_1 }} + z_1 \mathbf e_3 .
\end{equation}
Separating Eq.~(\ref{eq:threeunknownsC}) into its radial and axial components, we get
\begin{subequations} 
  \label{eq:threeunknownsCradax}
  \begin{align}
    \label{eq:threeunknownsCradial}
    s_1 \sin \theta_{ \sigma_1 } \, \mathbf e_1 e^{ i \mathbf e_3 \phi_{ \sigma_1 } } 
    + R \mathbf e_1 e^{i \mathbf e_3 \phi_{ \eta_0 }}
    &= 
    R \mathbf e_1 e^{i \mathbf e_3 \phi_{ \eta_1 }} ,
    \\
    \label{eq:threeunknownsCaxial}
    s_1 \cos \theta_{ \sigma_1 } \, \mathbf e_3 
    + z_0 \mathbf e_3 
    &= 
    z_1 \mathbf e_3 .
  \end{align}
\end{subequations}
Notice that there are three unknowns in Eq.~(\ref{eq:threeunknownsCradax}): $s_1$, $\phi_{ \eta_1 }$, $z_1$. 

To solve for the propagation distance $s_1$, we first square both sides of Eq.~(\ref{eq:threeunknownsCradial}):
\begin{equation}
  s_1^2 \sin^2 \theta_{ \sigma_1 }   
  +
  R^2
  + 
  2 s_1 R \sin \theta_{ \sigma_1 } \cos(\phi_{ \eta_0 }-\phi_{ \sigma_1 })
  = 
  R^2 .
\end{equation}
Hence,
\begin{equation}
\label{eq:s1soln}
  s_1
  = 
  - \frac{2 R \cos(\phi_{ \eta_0 }-\phi_{ \sigma_1 })}{\sin \theta_{ \sigma_1 } } .
\end{equation}

To solve for $\phi_{ \eta_1 }$, we expand Eq.~(\ref{eq:threeunknownsCradial}) into its $\mathbf e_1$ and $\mathbf e_2$ components:
\begin{subequations}
  \label{eq:threeunknownsCradialcomponents}
    \begin{align}
    \label{eq:threeunknownsCradialcomponents1}
    R \cos \phi_{ \eta_1 } &= s_1 \sin \theta_{ \sigma_1 } \cos \phi_{ \sigma_1 } + R \cos \phi_{ \eta_0 } ,
    \\
    \label{eq:threeunknownsCradialcomponents2}
    R \sin \phi_{ \eta_1 } &= s_1 \sin \theta_{ \sigma_1 } \sin \phi_{ \sigma_1 } + R \sin \phi_{ \eta_0 } .
  \end{align}
\end{subequations}
Dividing the two equations and using the expression for $s_1$ in Eq.~(\ref{eq:s1soln}), we get
\begin{equation}
  \label{eq:thanphieta2short}
\tan \phi_{ \eta_1 }
= \dfrac{ \sin \phi_{ \eta_0 }\cos 2\phi_{ \sigma_1 } - \cos \phi_{ \eta_0 } \sin 2\phi_{ \sigma_1 } }
{- \cos \phi_{ \eta_0 } \cos 2\phi_{ \sigma_1 }  - \sin \phi_{ \eta_0 } \sin 2\phi_{ \sigma_1 } }  ,
\end{equation}
after using the double angle identities.
Hence, 
\begin{equation}
\label{eq:phieta2}
\tan \phi_{ \eta_1 } = 
\dfrac{- \sin ( 2\phi_{ \sigma_1 } - \phi_{ \eta_0 }) }{ - \cos ( 2\phi_{ \sigma_1 } - \phi_{ \eta_0 })} .
\end{equation}
Note that we retained the negative signs in both the numerator and denominator so that we can properly determine the angle $\phi_{ \eta_1 }$ which is from $0$ to $2\pi$.


\section{Skew Ray Invariants}
\label{sec:invariants}

\subsection{Propagation Distance Between Reflections}

\begin{figure}
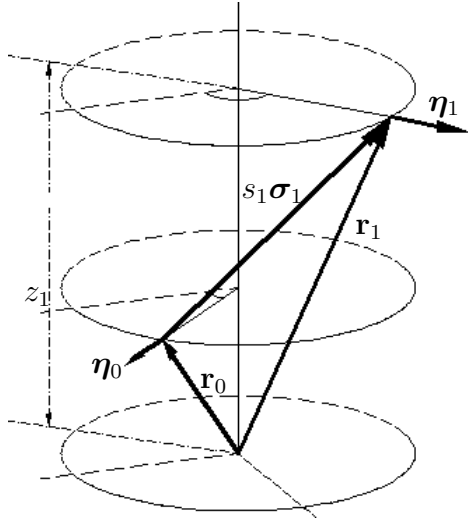

\centerline{
   \begin{overpic}[width=.7\columnwidth,tics=5,clip=true,page=7,trim=0mm 30.15mm 0mm 30.15mm]{skewrays_graphics}
     \put (3,42) {\large$z_1$}
     \put (16,28) {\large$\bm  \eta_0$}
     \put (81,79) {\large$\bm  \eta_1$}
     \put (37,25) {\large$\mathbf  r_0$}
     \put (67,55) {\large$\mathbf  r_1$}
     \put (45,62) {\large$s_1\bm \sigma_1$}
  \end{overpic}
}
   \caption{The light ray traveled from point $\mathbf r_0$ to point $\mathbf r_1$ on the fiber's walls, at a distance $s_1$, along the direction $\bm \sigma_1$. The normal vectors at the two points are $\bm \eta_0$ and $\bm \eta_1$.}
 \label{fig:secondprop}
\end{figure}

The procedure for computing the propagation of the light ray from $P_1$ to $P_2$ is the same as the procedure from $P_0$ to $P_1$. Therefore, we can simply change the subscripts to obtain the following relations for the propagation distance $s_2$, the propagation polar angle $\theta_{ \sigma_2 }$, and the propagation azimuthal angle $\phi_{\sigma2}$:
\begin{subequations}
\begin{align}
  \label{eq:s2soln}
  s_2
  &= 
  - \frac{2 R \cos(\phi_{ \eta_1 }-\phi_{ \sigma_2 })}{\sin \theta_{ \sigma_2 }} ,
\\
  \label{eq:thetasig2}
  \theta_{ \sigma_2 }
  &= \theta_{ \sigma_1 } ,
\\
\label{eq:phisig2}
  \phi_{\sigma2} 
  &= 2\phi_{ \eta_1 } - \phi_{ \sigma_1 } \pm \pi .
\end{align}
\end{subequations}
Notice that the angle $\theta_{\sigma}$ remains constant through reflections. 
Our aim is to show that the propagation distance $s$ is also constant between 
reflections, i.e., $s_1=s_2$. 

If we substitute Eqs.~(\ref{eq:thetasig2}) and (\ref{eq:phisig2}) into Eq.~(\ref{eq:s2soln}), we get
\begin{equation}
\label{eq:s2notnegative}
  s_2 = \dfrac{2R \cos (- \phi_{ \eta_1 } + \phi_{ \sigma_1 }) }{\sin \theta_{ \sigma_1 }} ,
\end{equation}
since $\cos(\phi-\pi)=-\cos \phi$.
Expanding Eq.~(\ref{eq:s2notnegative}) using a trigonometric identity,
and using the expression for the sines and cosines of $\phi_{ \eta_1 }$ in Eq.~(\ref{eq:phieta2}), 
\begin{align}
\sin \phi_{ \eta_1 } &= 
- \sin ( 2\phi_{ \sigma_1 } - \phi_{ \eta_0 })  ,
\\
\cos \phi_{ \eta_1 } &= 
 - \cos ( 2\phi_{ \sigma_1 } - \phi_{ \eta_0 }) ,
\end{align}
we obtain
\begin{equation}
\label{eq:s2equals1}
  s_2 = - \dfrac{2R \cos ( \phi_{ \eta_0 } - \phi_{ \sigma_1 }) }{\sin (\theta_{ \sigma_1 })} .
\end{equation}
Comparing this with the expression for $s_1$ in Eq.~(\ref{eq:s1soln}), we arrive at
\begin{equation}
\label{eq:s1iss2}
s_1 = s_2 .
\end{equation}
Thus, the propagation distance between two reflections is constant.

\subsection{Axial Distance Between Reflections}

The axial position $z_1$ of point $P_1$ is given 
in Eq.~(\ref{eq:threeunknownsCaxial}):
\begin{equation}
\label{eq:z2}
z_1
=
s_1 \cos \theta_{ \sigma_1 } + z_0 .
\end{equation}
Similarly, we can show that for point $P_2$,
\begin{equation}
\label{eq:z3}
    z_2
=
    s_2 \cos \theta_{ \sigma_2 }
    + z_1 .
\end{equation}
Taking the difference of Eq.~(\ref{eq:z2}) and Eq.~(\ref{eq:z3}), and imposing $s_2=s_1$ in Eq.~(\ref{eq:s2equals1}) and $\theta_{ \sigma_2 } = \theta_{ \sigma_1 } $ in Eq.~(\ref{eq:thetasig2}), then
\begin{equation}
\label{eq:deltaz}
\Delta z =
    z_2 - z_1
 = z_1-z_0 .
\end{equation}
Thus, the axial distance $\Delta z$ between wall-to-wall propagations, defined as the pitch, is constant.

If we substitute the expression for $z_1$ in Eq.~(\ref{eq:z2}) and $s_1$ in Eq.~(\ref{eq:s1soln}), we get
\begin{equation}
\Delta z = 
  - 2 R \cos(\phi_{ \eta_0 }-\phi_{ \sigma_1 }) \cot \theta_{ \sigma_1 } ,
\end{equation}
which is similar in form to that given in Eq.~(22) of Cozannet and Treheux\cite{Cozannet}.
Alternatively, we can express $\Delta z$ in terms of $s_1$ in Eq.~(\ref{eq:s1soln}) to get
\begin{equation}
\Delta z =
    s_1 \cos \theta_{ \sigma_1 } ,
\end{equation}
which is an invariant given in Love and Snyder\cite{lovesnyder}. Note that this invariant reduces only to the ray half-period if the rays are meridional.

\subsection{Change in Azimuthal Angle Between Reflections}

We know from Eq.~(\ref{eq:phieta2}) that the azimuthal angle $\phi_{ \eta_1 }$ at point $P_1$ is given by
\begin{equation}
\phi_{ \eta_1 } = \tan^{-1} \left( \dfrac{- \sin ( 2\phi_{ \sigma_1 } - \phi_{ \eta_0 }) }{ - \cos ( 2\phi_{ \sigma_1 } - \phi_{ \eta_0 })} \right ) ,
\end{equation}
so that
\begin{equation}
\label{eq:phieta2k}
\phi_{ \eta_1 } = 2\phi_{ \sigma_1 } - \phi_{ \eta_0 } + k \pi ,
\end{equation}
where $k$ is $+1$, $-1$, or $0$.

To find $k$, we substitute the expression for $\phi_{ \eta_1 }$ in Eq.~(\ref{eq:phieta2k}), for $\phi_{ \sigma_2 }$ in Eq.~(\ref{eq:phisig2}), and $\sin (\theta_{ \sigma_1 })$ in Eq.~(\ref{eq:thetasig2}) to Eq.~(\ref{eq:s2soln}) to obtain
\begin{equation}
  s_2
  = 
  - \frac{2 R \cos(2\phi_{ \sigma_1 } - \phi_{ \eta_0 } -2\phi_{ \eta_1 } + \phi_{ \sigma_1 } + (k \mp 1) \pi )}
{\sin \theta_{ \sigma_1 }} .
\end{equation}
Substituting the expression for $\phi_{ \eta_1 }$ in Eq.~(\ref{eq:phieta2k}) again and rearranging the terms, we get
\begin{equation}
  s_2
  = 
  - \frac{2 R \cos(-\phi_{ \sigma_1 } + \phi_{ \eta_0 } + (-k \mp 1) \pi )}
{\sin \theta_{ \sigma_1 }} .
\end{equation}
Comparing this the expression for $s_2$ in Eq.~(\ref{eq:s2equals1}), we arrive at
$k= \mp 1$ ,
so that Eq.~(\ref{eq:phieta2k}) reduces to
\begin{equation}
\label{eq:phieta2soln}
\phi_{ \eta_1 } = 2\phi_{ \sigma_1 } - \phi_{ \eta_0 } \mp \pi .
\end{equation}

Using a similar process, we can show that the form of $\phi_{ \eta_2 }$ and $\phi_{ \eta_3 }$ are similar to $\phi_{ \eta_1 }$, 
\begin{subequations}
\begin{align}
\label{eq:phieta3}
\phi_{ \eta_2 } &= 2\phi_{ \sigma_2 } - \phi_{ \eta_1 } \mp \pi ,
\\
\label{eq:phieta4}
\phi_{ \eta_3 } &= 2\phi_{\sigma 3} - \phi_{ \eta_2 } \mp \pi ,
\end{align}
\end{subequations}
so that 
\begin{subequations}
\begin{align}
\label{eq:deltaeta3eta2}
\phi_{ \eta_2 } - \phi_{ \eta_1 } &=  2(\phi_{ \sigma_2 } - \phi_{ \eta_1 }) \mp \pi ,
\\
\label{eq:deltaeta4eta3}
\phi_{ \eta_3 } - \phi_{ \eta_2 } &= 2(\phi_{\sigma 3} - \phi_{ \eta_2 }) \mp \pi .
\end{align}
\end{subequations}

Replacing the subscripts of Eq.~(\ref{eq:phisig2}) to obtain an expression for $\phi_{\sigma_3}$,
\begin{equation}
\label{eq:phisig3}
  \phi_{\sigma_3} = 2\phi_{\eta_3} - \phi_{\sigma_2} \pm \pi ,
\end{equation}
and using the the expressions for $\phi_{\sigma 3}$ in Eq.~(\ref{eq:phisig3}) and $\phi_{ \eta_2 }$ in Eq.~(\ref{eq:phieta3}), Eq.~(\ref{eq:deltaeta4eta3}) becomes
\begin{equation}
\label{eq:deltaeta4eta3fin}
\phi_{ \eta_3 } - \phi_{ \eta_2 } = 2(\phi_{ \sigma_2 } - \phi_{ \eta_1 }) \mp \pi .
\end{equation}
Comparing Eq.~(\ref{eq:deltaeta3eta2}) with Eq.~(\ref{eq:deltaeta4eta3fin}), we arrive at 
\begin{equation}
\Delta \phi =
\phi_{ \eta_3 } - \phi_{ \eta_2 } = \phi_{ \eta_2 } - \phi_{ \eta_1 } .
\end{equation}
Thus, the change in the azimuthal angle $\Delta \phi$ between consecutive reflections is a constant.

So now we have two invariants between consecutive reflections: the change in azimuthal angle $\Delta \phi$ and the change in the axial propagation distance $\Delta z$. From this, we conclude that the rays inside a fiber trace a polygonal helix.



\section{Conclusions}
\label{sec:conclusion}

In this paper, we used Geometric Algebra to compute the paths of skew rays in a cylindrical, step-index optical fiber. To do this, we used the vector addition form for the law of propagation, the exponential of an imaginary vector form for the law of refraction, and the vector product form for the law of reflection. In addition, we used the spherical angles $\theta$ and $\phi$ to describe the directions of rays in space, but expressed in cylindrical coordinates in exponential form. We showed that the light rays inside the optical fiber trace a polygonal helical path characterized by three invariants between sucessive reflections: (1) the ray path distance, (2) the change in axial distances, and (3) the change in the azimuthal angles. We also showed that the numerical aperture for skew rays we obtained is the same as that of the literature.


To derive the ray tracing invariants, we did not use the reflection and refraction laws in Klein and Furtak\cite{kleitak}. Rather, we used two alternative techniques: (a) the Pauli Identity which expresses the geometric product of two vectors as a sum of their dot and  imaginary cross products and (b) the Euler's Theorem which expresses the exponential of an imaginary vector as a sum of the cosine of the magnitude of the vector and the product of the normalized imaginary vector with the sine of the vector's magnitude. These two theorems allow us to express vector rotations in exponential form, which enables us to take advantage of the addition or subtraction of exponential arguments of two rotated vectors in the derivation of the ray tracing invariants. 

Many of the equations for the invariants were already known before, except maybe for the change in the azimuthal angle between reflections. Also, we obtained a new expression for the numerical aperture, which allows the point of entry of light to be an arbitrary point in the fiber's entrance, and not limited to the center or the edge as given by previous authors. Finally, our use of standard notations for the angles $\theta$ and $\phi$ for cylindrical and spherical coordinates, differentiated only by subscripts would hopefully help in visualizing the complex geometry of skew ray tracing. 

In future papers, we shall extend our work on skew rays to ray tracing in circular, conical, and toroidal fibers, which may be step-index, multistep-index, or graded index. 

\section*{Acknowledgement}

Thanks to James Hernandez for checking the readability of the manuscript. This work was supported by Manila Observatory and Ateneo de Manila University.

\bibliography{skew_ver1Bib}


\end{document}